\documentclass{ws-mpla} 
 
\newcommand{\be}{\begin{eqnarray}} 
\newcommand{\ee}{\nonumber \end{eqnarray}} 
 
\newcommand{\su}{{su}} 
 
\newcommand{\Sl}{{sl}} 
 
\newcommand{\nn}{\nonumber} 
\newcommand{\METgamma}{{\tilde\gamma}} 
\newcommand{\Mgamma}{{\gamma}} 
\newcommand{\del}{\partial}

\def\ket#1{\left|#1\right>} 
 
\newcommand{\vc}{\bar v} 
 
\def\pint{{-\!\!\!\!\!\!\int}} 
\renewcommand{\Im}{\, {\rm Im}}

\begin{document} 
 
\markboth{Ian Swanson} 
{A Review of Integrable Deformations in AdS/CFT} 
 
\catchline{}{}{}{}{} 
 
\title{A REVIEW OF INTEGRABLE DEFORMATIONS IN ADS/CFT} 
 
\author{\footnotesize IAN SWANSON} 
 
\address{School of Natural Sciences, Institute for Advanced Study, Einstein Drive\\ 
Princeton, New Jersey 08540, 
USA \\ 
swanson@sns.ias.edu} 
 
%
 
\maketitle 

 
\begin{abstract} 
Marginal $\beta$ deformations of ${\cal N}=4$ super-Yang-Mills theory are known to
correspond to a certain class of deformations of the $S^5$ background subspace of type
IIB string theory in $AdS_5\times S^5$. An analogous set of deformations of the $AdS_5$
subspace is reviewed here. String energy spectra computed in the near-pp-wave limit of
these backgrounds match predictions encoded by discrete, asymptotic Bethe equations,
suggesting that the twisted string theory is classically integrable in this regime.
These Bethe equations can be derived algorithmically by relying on the existence of Lax
representations, and on the Riemann-Hilbert interpretation of the thermodynamic Bethe
ansatz. This letter is a review of a seminar given at the Institute for Advanced Study,
based on research completed in collaboration with McLoughlin.
\ \\
\ \\
\noindent
{\it
Electronic version of an article published as 
[Modern Physics Letters A, Vol. 22, No. 13 (2007) 915-930] 
[doi:10.1142/S0217732307023614] \copyright~ 
[copyright World Scientific Publishing Company] [http://www.worldscinet.com/mpla/mpla.shtml].}
\end{abstract} 
 
\ccode{PACS Nos.: 11.25.Tq, 11.25.Hf, 02.30.Ik} 
 
\section{Introduction} 
In recent years a large number of studies have emerged indicating 
that type IIB string theory on $AdS_5\times S^5$ and ${\cal N}=4$ 
super-Yang-Mills (SYM) theory in four dimensions may be integrable 
in the planar limit.  The technology of integrable systems has 
therefore become extraordinarily useful in studying the AdS/CFT 
correspondence in detail.  The correspondence equates the spectrum 
of operator anomalous dimensions of the gauge theory with the energy 
spectrum of the string theory. The rank of the Yang-Mills 
gauge group determines the amount of Ramond-Ramond (RR) flux on the 
$S^5$ subspace in the string theory, and in the planar limit this 
number is scaled to infinity: $N_c \to \infty$.  
The string coupling $g_s$ is related 
to the gauge theory coupling $g_{\rm YM}$ via the standard relation, 
$g_s = e^{\phi_0} = {g^2_{\rm YM}/ 4\pi}$, and the radial scale of 
both the $AdS_5$ and $S^5$ spaces is given by $R^4= 4 \pi g_s N_c = 
g^2_{\rm YM}N_c = \lambda$ (with $\alpha' = 1$).

If these theories are indeed integrable, the dynamics should be 
encoded in a diffractionless scattering matrix $S$.  On the string 
side, in the strong-coupling limit $(\lambda = g_{\rm YM}^2 N_c \to 
\infty)$, this $S$ matrix can be interpreted as describing the 
two-body scattering of elementary excitations on the worldsheet. 
As their worldsheet momenta becomes large, these 
excitations are better described as special types of solitonic 
solutions, or {\it giant magnons},  and the interpolating region is described by the 
dynamics of the so-called near-flat-space 
regime.\cite{giantmagnons,NFS}  On the gauge theory side, the action 
of the dilatation generator on single-trace operators can be equated 
with that of a Hamiltonian acting on states of a spin 
chain.\cite{MZ}  In this picture, operators in the trace are 
represented as lattice pseudoparticles that, like their stringy 
counterparts, experience diffractionless scattering encoded by an 
$S$ matrix.  Proving that the gauge and string theories are identical 
in the planar limit
therefore amounts to showing that the underlying physics of both 
theories is governed by the same two-body scattering matrix. In 
fact, symmetry fixes this $S$ matrix up to an overall phase 
$\sigma$, so what remains is to somehow determine $\sigma$ from 
basic principles.\cite{Beisertsu22}  (Unitarity and crossing 
relations, as they exist in this context, constrain this phase to 
some extent; see Refs.~\refcite{Janik:2006dc,Beisert:2006ib,Beisert:2006ez} for 
recent developments.)

An impressive amount of evidence exists in favor of the mutual 
integrability of these two theories.  If true, this raises the 
question of whether these theories can be deformed in a controlled 
manner while remaining integrable.  One class of interesting 
deformations to consider are the marginal $\beta$ deformations of 
${\cal N}=4$ SYM, also known as Leigh-Strassler 
deformations.\cite{LS}  The resulting theories comprise a 
one-parameter family of ${\cal N}=1$ conformal gauge theories (in 
the case of real $\beta$ deformations). On the gravity side of the 
correspondence, these correspond to special geometrical deformations 
of the $S^5$ subspace in the string theory background.\cite{LM} 
 
In fact, the integrability of the gauge and string theory, to the 
extent that it is understood in the undeformed cases, seems to 
persist under these deformations.  This problem was studied directly 
and indirectly, for example, in Refs.~\refcite{Frolov:2005ty,Plefka:2005bk,Frolov:2005dj,Alday:2005ww,Freyhult:2005ws,Chen:2005sb,Chen:2006bh,Beisert:2005if,Spradlin:2005sv,Bobev:2005cz,Ryang:2005pg} (see 
also references therein).  The dynamics of both theories can be 
captured, at least in certain limits, by {\it twisted} Bethe 
equations.  Here we review an analogous class of deformations acting 
on the $AdS_5$ subspace of the string theory background, first 
studied in Ref.~\refcite{US}. While the corresponding gauge theory 
is less well understood (it may be a non-commutative or even 
non-associative theory), the string theory seems to be well defined 
in the near-pp-wave regime.  Furthermore, the string energy spectrum 
can be computed precisely in this limit from a discrete Bethe 
ansatz, which lends substantial support to the methodology developed 
in Refs.~\refcite{Arutyunov:2004vx,Staudacher:2004tk,Beisert:2005fw}.

In Section \ref{DEFS} below, TsT deformations of the string 
background geometry are reviewed in detail.  The classical 
integrability of the string sigma model is discussed in Section 
\ref{INTEG}.  String energy spectra are computed directly from the 
deformed Green-Schwarz action in the near-pp-wave limit in Section 
\ref{SPEC}.  In Section \ref{BETHE}, the thermodynamic Bethe 
equations are promoted to discrete Bethe equations that correctly 
reproduce the deformed energy spectra.  A brief discussion and 
thoughts on further research are given in Section \ref{CONC}.  This 
letter is a review of a seminar based on Ref.~\refcite{US} given in 
May, 2006 at the Institute for Advanced Study.

\section{Deformations of the string background geometry} 
\label{DEFS} 
In Ref.~\refcite{LM}, it was demonstrated that marginal $\beta$ 
deformations of ${\cal N}=4$ SYM theory correspond to specific 
deformations of the $AdS_5\times S^5$ background geometry of the 
dual gravity theory.  In general, these transformations act on 
global toroidal (or $U(1)\times U(1)$) isometries of the string 
theory background, and have been dubbed $\gamma$ deformations. The 
generic action of these deformations can be understood as arising 
from the following sequence of transformations: 
\begin{romanlist}[(ii)] 
\item  T duality acting on the first $U(1)$ factor of the global symmetry; 
\item  a coordinate shift parameterized by $\gamma$ acting on the second 
    $U(1)$ factor; 
\item  T duality acting again on the first $U(1)$. 
\end{romanlist} 
For this reason, these $\gamma$ deformations are also known as {\it 
TsT deformations}. 
 
\subsection{Deformations on the $S^5$ subspace} 
In keeping with notation used in the literature, we will 
parameterize the coordinate shift with the symbols $\METgamma_i$. 
For the case of real $\METgamma_i$, the deformed spacetime metric 
and background fields are given by (mostly following the notation of 
Ref.~\refcite{Frolov:2005iq}, and with $\alpha'=1$), 
\begin{eqnarray} 
ds^2_{\rm string}/R^2 &=& ds^2_{{AdS_5}} + 
   \sum^3_{i=1} ( d\rho_i^2  + G \rho_i^2 d\phi_i^2) +   G 
\rho_1^2\rho_2^2\rho_3^2 [d (\sum^3_{i=1} \METgamma_i \phi_i)]^2\ , 
\nn\\ 
&& 
\nn\\ 
B_2 &=&  R^2  G w_2 \ , 
    \qquad 
    e^\phi = e^{\phi_0}G^{1/2}\ , 
\nn\\ 
&& 
\nn\\ 
w_2  &\equiv&  \METgamma_3\,  \rho_1^2 \rho_2^2\, d\phi_1 \wedge d\phi_2 + 
\METgamma_1\,  \rho_2^2 \rho_3^2\, d\phi_2 \wedge d\phi_3 + \METgamma_2\, 
\rho_3^2 \rho_1^2\, d\phi_3 \wedge d\phi_1\ ,  \nn\\ 
&& 
\nn\\ 
G^{-1} &\equiv& 1 +  \METgamma_3^2\,  \rho_1^2 \rho_2^2 +  \METgamma_1^2\, 
\rho_2^2 \rho_3^2 +  \METgamma_2^2\,  \rho_1^2 \rho_3^2\ . 
\nn 
\end{eqnarray} 
Here, the $S^5$ subspace has undergone three consecutive TsT 
deformations parameterized by $\METgamma_i$, with $i \in 1,2,3$. 
$B_2$ is the NS-NS two-form field strength, and the two- and 
five-form field strengths $C_2$ and $F_5$ have been omitted.  The 
angular coordinates on the sphere can be written as \be 
\rho_1 = \sin \alpha \cos \theta\ , \ \ \ 
\rho_2 =  \sin \alpha \sin \theta\ , \ \ \ 
 \rho_3=  \cos \alpha\ , 
\ee  
such that $\sum_{i=1}^3\rho_i^2=1$. 
 
We find it convenient to use the following lightcone parametrization 
on $S^5$: \be 
&&    \rho_2  = \frac{y_1}{R}~, \qquad 
    \rho_3  =  \frac{y_2}{R}\ , \qquad    \rho_3  = \sqrt{1-\rho_2^2-\rho_3^2}\ , 
    \nn\\ 
    && 
    \phi_1  = x^+ +\frac{x^-}{R^2}\ , \qquad 
    t = x^+\ . 
\ee 
At large $R$, we reach a semiclassical limit described by 
point-like (or ``BMN'') strings boosted to lightlike momentum $J$ 
along a geodesic on the deformed $S^5$ 
subspace.\cite{Berenstein:2002jq} The angular momentum $J$ in the 
$\phi_1$ direction is related to the scale radius $R$ by $p_- R^2 = 
J$, and the lightcone momenta take the form \be  -p_+ = \Delta-J~, 
\qquad -p_-=i\partial_{x^-} = \frac{i}{R^2}\partial_{\phi_1} 
    = -\frac{J}{R^2}\ . 
\ee 
We also find it convenient to work with the following form of 
the spacetime metric on $AdS_5$: 
\be ds^2_{AdS_5}=-\left(\frac{1+ 
x^2/4 R^2}{1- x^2/4 R^2}\right)^2 dt^2 
    + \frac{dx^2/R^2}{(1-x^2/4 R^2)^2}\ , 
\ee  
where $x$ is a vector spanning an $SO(4)$ subspace transverse to 
the lightcone. 
 
To simplify the projection onto closed $su(2)$ subsectors of the 
theory, we introduce the following complex coordinates: \be 
y&=& y_1 \cos \phi_2+i y_1 \sin \phi_1\ , \qquad 
    \bar{y}=y_1 \cos \phi_2 - i y_1 \sin \phi_1\ ,\nn\\ 
z&=& y_2 \cos \phi_2+i y_2 \sin \phi_1\ , \qquad 
    \bar{z}=y_2 \cos \phi_2 - i y_2 \sin \phi_1\ . 
\ee  
Defining the large-$R$ expansion of the spacetime metric by  
\be 
ds^2 = ds_{(0)}^2 + \frac{ds_{(1)}^2}{R^2} + O(1/R^4)\ ,  
\ee  
we obtain  
\be 
ds_{(0)}^2&=&2dx^+dx^- + |dy|^2 + |dz|^2 -(dx^+)^2 
    \left[x^2+|y|^2(1+\METgamma_3^2)+|z|^2(1+\METgamma^2_2)\right]\ , 
\nn\\ 
ds_{(1)}^2 &=& 
    (dx^-)^2+\frac{1}{4}(yd\bar{y}+\bar{y}dy+zd\bar{z}+\bar{z}dz)^2 
    +\frac{1}{2}x^2dx^2 
\nn\\ 
& & \kern-18pt 
    +(dx^+)^2\Bigl[ -\frac{1}{2}x^4 +2(|z|^2+|y|^2)(|z|^2\METgamma_2^2 
    +|y|^2\METgamma_3^2)+(|y|^2\METgamma_3^2+|z|^2\METgamma_2^2)^2  \Bigr] 
\nn\\ 
& &\kern-18pt+\METgamma_1 dx^+(\METgamma_2 |z|^2 
\Im(\bar{y}dy)+\METgamma_3 |y|^2 
    \Im(\bar{z}dz))-\METgamma_3^2\Im(\bar{y}dy)^2 
    -\METgamma_3^2\Im(\bar{z}dz)^2 
\nn\\ 
& &\kern-18pt+ \METgamma_2\METgamma_3\Im(\bar{y}dy)\Im(\bar{z}dz) 
-2dx^+ dx^-(|y|^2(1+\METgamma_3^2)+|z|^2(1+\METgamma_2^2))\ . \ee As 
expected, the pp-wave metric appears at leading order in the 
expansion. The corresponding expansion of the NS-NS two-form $B_2$ 
appears as \be 
B_2&=&\METgamma_3 dx^+\wedge \Im (\bar{y}dy)-\METgamma_2 dx^+ \wedge \Im(\bar{z}dz)\nn\\ 
& &\kern-15pt +\frac{1}{R^2}\left[ 
-\METgamma_3(\METgamma_3^2|y|^2+\METgamma_2^2|z|^2)dx^+\wedge 
\Im(\bar{y}dy)+\METgamma_2(\METgamma_3^2|y|^2 
    +\METgamma_2^2|z|^2)dx^+\wedge\Im(\bar{z}dz)\right.\nn\\ 
& &\kern-15pt \left. +\METgamma_3 dx^- 
  \Im(\bar{y}dy)-\METgamma_2dx^-\wedge\Im(\bar{z}dz)+\METgamma_1\Im(\bar{y}dy)\wedge\Im(\bar{z}dz)\right]\ . 
\ee

The rank-one $su(2)$ subsector decouples from the theory in the 
near-pp-wave limit, though it has nontrivial dynamics itself.  We 
can therefore study is as a separate theory, without having to worry 
about mixing.  The TsT-deformed version, labeled by $su(2)_\gamma$, 
can be reached by projecting onto a single complex coordinate 
(either $y$ or $z$ for the parametrization given above): \be 
ds^2_{\su(2)_\gamma}&=&2 dx^+dx^--\left(1+\METgamma^2\right)|y|^2 
(dx^+)^2+|dy|^2+\frac{1}{R^2} 
    \Big[ \frac{1}{4}(yd{\bar y} 
    +{\bar y}dy)^2 
\nn\\ 
&&\kern-20pt 
    +(dx^-)^2+\METgamma^2(2+\METgamma^2)|y|^4(dx^+)^2-2(1+\METgamma^2)|y|^2dx^+dx^- 
        -\METgamma \Im\left({\bar y}dy\right)\Big] \ , 
\ee where we have truncated the series at $O(1/R^4)$. Here we 
allow the parameter $\METgamma$ to stand for either $\METgamma_2$ or 
$\METgamma_3$, corresponding to two possible choices of 
$\su(2)_\gamma$ truncation. The corresponding NS-NS two-form reduces 
to \be B_2^{\su(2)_\gamma}&=&\METgamma dx^+ \wedge \Im({\bar 
y}dy)+\frac{\METgamma}{R^2}\Bigl(dx^-\wedge \Im({\bar y}dy) 
\nn\\ 
&& 
    -(1+\METgamma^2)|y|^2 dx^+ \wedge \Im({\bar y}dy)\Bigr)+ O(1/R^4)\ . 
\ee

\subsection{Deformations in the $AdS_5$ subspace} 
To study analogous deformations of the $AdS_5$ subspace, it is 
convenient to start with an $SO(4,2)$ invariant, expressed in terms 
of ${\mathbb R}^6$ embedding coordinates: 
\be 
 -X_0^2 
+ X_1^2 + X_2^2+ X_3^2+ X_4^2- X_5^2 = -1\ , \ee such that \be 
X_0 = \eta_1 \sin\hat\varphi_1\ , &\qquad & X_1 = \eta_2\cos\hat\varphi_2\ , \nn\\ 
X_2 = \eta_2 \sin\hat\varphi_2\ , &\qquad & X_3 = \eta_3\cos\hat\varphi_3\ , \nn\\ 
X_4 = \eta_3 \sin\hat\varphi_3\ , &\qquad & X_5 = 
\eta_1\cos\hat\varphi_1\ . \ee The $\hat\varphi_i$ variables denote 
untwisted $U(1)$ angular coordinates.  One can make contact with the 
more familiar angular coordinates on $AdS_5$ using 
\be \eta_1 = 
\cosh\alpha\ , 
    \qquad 
    \eta_2 = \sinh\alpha \sin\theta\ , 
    \qquad 
    \eta_3 = \sinh\alpha\cos\theta\ . 
\ee This preserves the $SO(2,1)$ invariant $-\eta_1^2 + \eta_2^2 + 
\eta_3^2 = -1$. 
 From the 
resulting metric \be {ds_{AdS_5}^2}/{R^2} 
     & = & 
    -(d\eta_1^2 + \eta_1^2 d\hat\varphi_1^2) 
    + \sum_{i=2}^3 (d\eta_i^2 + \eta_i^2 d\hat\varphi_i^2) 
\nn\\ 
& = &    d\alpha^2 - \cosh\alpha^2 d\hat\varphi_1^2 
    + \sinh\alpha^2 \left( 
    d\theta^2 + \sin\theta^2 d\hat\varphi_2^2 
    + \cos\theta^2d\hat\varphi_3^2 \right)\ , 
\ee it is easy to see that there is a $U(1)\times U(1)\times U(1)$ 
global symmetry.  TsT deformations may thus act on the corresponding 
angular coordinates $\hat\varphi_i$ ($i\in 1,2,3$).  It turns out 
that we can avoid the usual difficulties associated with T-duality 
along compact timelike directions (see Ref.~\refcite{US} for 
details). In the end, however, one must pass to the universal 
covering space of the geometry, where the timelike directions are 
noncompact.

For present purposes, we wish to study a TsT transformation that 
acts as a T-duality along the $\hat\varphi_2$ direction, followed by 
a shift in the $\hat\varphi_1$ direction $\hat\varphi_1 \to 
\hat\varphi_1 + \METgamma \hat\varphi_2$, and a final T-duality 
along the new $\varphi_2$ direction (where the notation $\varphi_2$ 
indicates a transformed angular coordinate).   The deformed metric 
thus takes the form \be 
ds^2_{\rm str}/R^2 =  ds^2_{_{S^5}} + 
      g^{ij} d\eta_i d\eta_j 
        + g^{ij}\, G\, \eta_i^2 d\varphi_j^2 
    - \METgamma^2\, G\,  \eta_1^2\eta_2^2\eta_3^2\,  d  \varphi_1^2 \ , 
\ee with $g = {\rm diag}(-1,1,1)$.  The NS-NS two-form is \be B_2 = 
\METgamma\, G\, \eta_1^2 \eta_2^2\, d\varphi_1 \wedge d\varphi_2\ , 
\ee where the deformation factor $G$ is given by \be G^{-1} &\equiv& 
1 - \METgamma^2  \eta_1^2 \eta_2^2\ . \ee We focus on the local 
region for which $\varphi_2$ is spacelike:  in the end this yields a 
worldsheet Hamiltonian that appears to be completely consistent.

The $AdS_5^\METgamma$ worldsheet action appears as 
\be 
S_{{AdS_5}^{\METgamma}} & = & -\frac{\sqrt{\lambda}}{2} 
    \int d\tau \frac{d\sigma}{2\pi} 
    \Bigl[ \gamma^{\alpha\beta} 
    \left( 
    g^{ij} \del_\alpha\eta_i \del_\beta\eta_j 
    + g^{ij}\, G\,  \eta_i^2 \del_\alpha \varphi_j \del_\beta\varphi_j 
    \right. 
    \nn\\ 
    && 
    \kern-25pt 
    \left. 
    - \METgamma^2\, G\,  \eta_1^2\eta_2^2\eta_3^2\, 
    \del_\alpha\varphi_1\del_\beta\varphi_1 
    \right) 
    -2 \epsilon^{\alpha\beta}\left( 
    \METgamma\, G\, \eta_1^2\eta_2^2\, 
    \del_\alpha \varphi_1 \del_\beta \varphi_2 
    + \Lambda (g^{ij}\eta_i\eta_j +1 ) 
    \right)\Bigr]\ , 
\ee where $\Lambda$ acts as a Lagrange multiplier enforcing 
$-\eta_1^2 + \eta_2^2 + \eta_3^2 = -1$ on shell.  The indices 
$\alpha$ and $\beta$ run over the $\tau$ ($\alpha,\ \beta = 0$) and 
$\sigma$ ($\alpha,\ \beta = 1$) directions on the worldsheet, and 
$\gamma^{\alpha\beta}$ is the worldsheet metric.

To simplify the projection onto closed $sl(2)_\gamma$ subsectors 
(i.e.,~the analogues of the closed $su(2)_\gamma$ subsectors in the 
$S^5$ case), we choose the lightcone parametrization  
\be  
&& \eta_2 = 
\frac{u_1}{R}~, \qquad 
    \eta_3  =  \frac{u_2}{R}\ , \nn 
    \\ 
&&    \eta_3  = \sqrt{1+\eta_2^2+\eta_3^2}\ , \qquad 
    \varphi_1  = x^+ +\frac{x^-}{R^2}\ , \qquad 
    t = x^+\ , 
\ee and introduce a new set of complex coordinates: \be 
v&=& u_1 \cos \varphi_2+i u_1 \sin \varphi_1\ , \qquad  \bar{v}=u_1 \cos \varphi_2 
    - i u_1 \sin \varphi_1\ ,\nn\\ 
w&=& u_2 \cos \varphi_2+i u_2 \sin \varphi_1\ , \qquad  \bar{w}=u_2 \cos \varphi_2 
    - i u_2 \sin \varphi_1\ . 
\ee Under one such projection, the metric and NS-NS two-form become: 
\be ds^2_{\Sl(2)_\gamma} & = & 2 dx^+ dx^- -  (1+\METgamma^2)|v|^2 
(dx^+)^2 + |dv|^2 
    - \frac{1}{R^2}\Bigl[ 
    \frac{1}{4}(v d{\bar v} 
    +{\bar v}dv)^2 - (dx^-)^2 
\nn\\ 
&&  + \METgamma^2(2+\METgamma^2)|v|^4(dx^+)^2 
        + \METgamma^2 (v d\bar v - \bar v dv)^2 \Bigr] + O(1/R^4)\ , 
\nn\\ 
B_2^{\Sl(2)_\gamma} &=& \frac{i}{2}\METgamma dx^+ \wedge (v d\bar v - \bar v dv) 
    +\frac{i}{2R^2} |v|^2 \METgamma (1+\METgamma^2) dx^+ \wedge 
    (\bar v dv - v d\bar v)+ O(1/R^4)\ . 
\ee

The conserved $U(1)$ currents $J_i^\alpha$ in the undeformed theory 
turn out to be identical to those in the deformed theory. By 
defining canonical momenta as $p_i = J_i^0$, the associated charges 
take the form $J_i = \int \frac{d\sigma}{2\pi} p_i$.  The 
identification $\hat J_i = J_i$ therefore yields \be 
\hat\varphi_1' =    \varphi_1' - \Mgamma\, p_2\ , \qquad 
\hat\varphi_2' =    \varphi_2' + \Mgamma\, p_1\ , \qquad 
\hat\varphi_3' =    \varphi_3'\ . \ee where $\varphi'$ denotes a 
worldsheet $\sigma $ derivative acting on $\varphi$, and we have 
introduced the rescaled deformation parameter $ \Mgamma \equiv 
\frac{\METgamma}{\sqrt{\lambda}}$. We therefore obtain the following 
set of twisted boundary conditions \be 
\hat\varphi_1(2\pi) - \hat\varphi_1(0) & = & 2\pi(m_1 -\Mgamma\, J_2)\ , \nn\\ 
\hat\varphi_2(2\pi) - \hat\varphi_2(0) & = & 2\pi(m_2 +\Mgamma\, J_1)\ , \nn\\ 
\hat\varphi_3(2\pi) - \hat\varphi_3(0) & = & 2\pi m_3\ , \ee where 
the $m_i$ are winding numbers, satisfying $2\pi m_i = 
\varphi_i(2\pi) - \varphi_i(0)$.

\section{Classical integrability} 
\label{INTEG} 
We will now review how the classical integrability of the theory on 
the $AdS_5$ subspace is preserved under the deformation described 
above, relying on a parametrization of the bosonic coset space 
$(SO(4,2)\times SO(6)) / (SO(5,1)\times SO(5))$ originally given in 
Ref.~\refcite{Arutyunov:2004yx}.  The $AdS_5$ sector takes the form 
\be g = \left( 
\begin{array}{cccc} 
0       & Z_1       & -Z_3      & \bar Z_2  \\ 
-Z_1        & 0     & Z_2       & \bar Z_3  \\ 
Z_3     & -Z_2      & 0     & -\bar Z_1 \\ 
-\bar Z_2   & -\bar Z_3 & \bar Z_1  & 0 
\end{array} 
\right)\ , \qquad Z_i \equiv \eta_i e^{i \hat\varphi_i}\ , 
\ee 
which satisfies 
\be 
g^\dag\, s\, g = s\ , \qquad s \equiv {\rm diag}(-1,-1,1,1)\ . 
\ee 
The result is that $g$ is an $SU(2,2)$ embedding of an element of the coset 
$SO(4,2)/SO(5,1)$.   We can therefore work from the principal chiral model 
defined by 
\be 
S = \int d\tau d\sigma \gamma^{\alpha\beta} 
    {\rm Tr} \left( g^{-1} \del_\alpha g\, g^{-1} \del_\beta g\right)\ . 
\ee

Key to the analysis is the existence of a Lax representation, which 
encodes the equations of motion $\del_\alpha 
(\gamma^{\alpha\beta}j_\beta) = 0$ in an auxiliary linear problem, 
subject to a constraint equation in the form of a commutator of Lax 
operators $\left[ D_\alpha, D_\beta \right] = 0$. In the case at 
hand, the Lax operator $D_\alpha$ can be defined in terms of a 
spectral parameter $x$ by \be 
D_\alpha = \del_\alpha - \frac{j_\alpha^+}{2(x-1)} + \frac{j_\alpha^-}{2(x+1)} 
    \equiv \del_\alpha - {\cal A}_\alpha(x)\ . 
\ee ${\cal A}_\alpha (x)$ is the right Lax connection, and the 
$j_\alpha^+$ and $j_\alpha^-$ are respectively self-dual and 
anti-self-dual projections of the right current $j_\alpha = g^{-1} 
\del_\alpha g$. 
It turns out that the non-derivative dependence of the Lax current 
$j_\alpha$ on the angular coordinates $\hat\varphi_i$ can be gauged away 
using an invertible matrix $M$ 
\be 
\tilde\jmath_\alpha(\eta_i,\del\hat\varphi_i) &=& 
    M j_\alpha(\eta_i,\hat\varphi_i) M^{-1} 
\nn\\ 
     &=& \tilde g^{-1} \del_\alpha \tilde g 
    + \tilde g^{-1} \del_\alpha \Phi \tilde g 
    + \del_\alpha \Phi \ . 
\ee 
 
We now truncate to the deformed $sl(2)_\gamma$ sector of the theory. 
Strings in this subsector propagate on $AdS_3\times S^1$, though the 
$S^1$ factor will decouple. A useful coordinate parametrization can 
be found using the following $SL(2)$ matrix: \be 
g = \left( 
\begin{array}{cc} 
\cos\hat\varphi_1 \cosh\rho + \cos\hat\varphi_2 \sinh\rho & 
        \sin\hat\varphi_1 \cosh\rho - \sin\hat\varphi_2 \sinh\rho \\ 
-\sin\hat\varphi_1 \cosh\rho - \sin\hat\varphi_2 \sinh\rho & 
        \cos\hat\varphi_1 \cosh\rho - \cos\hat\varphi_2 \sinh\rho 
\end{array} 
\right)\ . 
\ee 
In this case we can invoke a gauge transformation of the form 
\be 
g = e^{\frac{i}{2}(\hat\varphi_1+\hat\varphi_2)\sigma_2} 
    e^{\rho\sigma_3} e^{\frac{i}{2}(\hat\varphi_1-\hat\varphi_2)\sigma_2}\ , 
\ee which eliminates any linear dependence on the coordinates 
$\hat\varphi_i$ ($\sigma_i$ are the usual Pauli matrices).  With $M 
= e^{\frac{i}{2}(\hat\varphi_1-\hat\varphi_2)\sigma_3}$, the right 
current takes the form \be \tilde 
\jmath_\alpha(\eta_i,\del\hat\varphi_i)  & =& M j_\alpha 
(\eta_i,\hat\varphi_i) M^{-1} 
\nn\\ 
    &&\kern-30pt =  \left( 
\begin{array}{cc} 
\del_\alpha \rho & e^{-\rho} ( \del_\alpha\hat\varphi_1\cosh\rho - 
\del_\alpha\hat\varphi_2\sinh\rho ) \\ 
- e^{-\rho} ( \del_\alpha\hat\varphi_1\cosh\rho + \del_\alpha\hat\varphi_2\sinh\rho ) & 
-\del_\alpha\rho 
\end{array} 
\right)\ . \ee We thus find the following local Lax operator and 
associated Lax connection: \be D_\alpha &\to&  M D_\alpha M^{-1} 
\equiv \del_\alpha - {\cal R}_\alpha\ , \nn 
\\ 
{\cal R}_\alpha  &=&  M {\cal A}_\alpha M^{-1} - M \del_\alpha M^{-1} 
    = \tilde {\cal A}_\alpha + \frac{i}{2}(\del_\alpha\hat\varphi_1 
    - \del_\alpha\hat\varphi_2)\sigma_2\ . 
\ee

Thermodynamic Bethe equations can be derived to encode the spectral 
problem by studying the pole structure and the asymptotics of the 
quasimomentum $p(x)$ on the complex spectral $x$-plane. The 
quasimomentum $p(x)$ is defined in the usual fashion, in terms of a 
monodromy $\Omega(x)$, according to \be {\rm Tr}\, \Omega(x) = 2\cos 
p(x)\ , \ee where \be \Omega(x) = {\cal P} \exp \int_0^{2\pi} 
d\sigma\, {\cal R}_1(x)\ . \ee The general strategy is to 
reformulate the Bethe ansatz as a Riemann-Hilbert 
problem.\cite{Kazakov:2004qf} The gauge freedom noted above turns 
out to be advantageous when applying these techniques in the 
presence of $\gamma$ deformations. In fact, we find that the poles 
of the quasimomentum at $x = \pm 1$ are {\it invariant} under 
$\gamma$ deformations: \be p(x) = \pi \frac{J/\sqrt{\lambda} \mp 
m}{x\pm 1} + \cdots \qquad x \to \mp 1\ , \ee where $m$ indicates a 
winding number associated with the decoupled $S^1$ in the 
$AdS_3\times S^1$ subspace. 
 
Following the treatment in Ref.~\refcite{Frolov:2005ty}, we find 
that it is easiest to study the asymptotics of the problem by using 
an inverse gauge transformation and relying on the original Lax 
connection ${\cal A}_\alpha$: \be T(x) = M(2\pi) {\cal P} \exp 
\int_0^{2\pi} d\sigma {\cal A}_1(x) M^{-1}(0)\ . \ee This gives a 
representation of $p(x)$ of the form \be 
2\cos p(x) &=& {\rm Tr}\ M_R\, {\cal P} \exp \int_0^{2\pi} d\sigma {\cal A}_1(x)\ , 
\nn\\ 
{\cal A}_1(x) & = & \frac{j_1}{x^2-1} + \frac{x\, j_0}{x^2-1}\ ,\ee 
where \be M_R = M^{-1}(0) M(2\pi) = \left( 
\begin{array}{cc} 
\cos  \Mgamma\pi (S-\Delta) & -\sin \Mgamma\pi (S-\Delta) \\ 
\sin  \Mgamma\pi (S-\Delta) & \cos \Mgamma\pi (S-\Delta) 
\end{array} 
\right)\ . \ee Here we have defined $J_1 = -\Delta$ and $J_2 = S$ to 
make contact with the energy $\Delta$ and impurity number $S$.

Using the right and left currents \be j_\alpha = g^{-1} \del_\alpha 
g = \frac{1}{2}j_\alpha \cdot \hat\sigma\ , \qquad l_\alpha = 
\del_\alpha g g^{-1} = \frac{1}{2}l_\alpha \cdot \hat\sigma\ , \ee 
with $\hat\sigma \equiv (i\sigma_2,\sigma_3,-\sigma_1)$, we find \be 
\frac{\sqrt{\lambda}}{4\pi}\int_0^{2\pi}d\sigma\, j_0^0 = \Delta +S\ 
, \qquad \frac{\sqrt{\lambda}}{4\pi}\int_0^{2\pi}d\sigma\, l_0^0 = 
\Delta -S\ . \ee At this point we may simply expand $p(x)$ in the 
asymptotic regions and, following the prescription described in 
Ref.~\refcite{Frolov:2005ty}, discard nonlocal contributions to 
recover \be 
p(x) &=& \pi\Mgamma (\Delta-S) + 2\pi \frac{\Delta+S}{\sqrt{\lambda}\, x} + \cdots\ , \qquad 
    x \to \infty\ , 
\nn\\ 
p(x) &=& \pi\Mgamma (\Delta+ S) - 2\pi \frac{\Delta-S}{\sqrt{\lambda}} x + \cdots\ , \qquad 
    x \to 0\ . 
\ee With the above input, one may now define a resolvent: \be G(x) = 
p(x) - \pi\frac{J/\sqrt{\lambda}+m}{x-1} - 
\pi\frac{J/\sqrt{\lambda}-m}{x+1} 
    - \pi\Mgamma (\Delta - S)\ . 
\ee The asymptotics of $G(x)$ are completely determined by the 
corresponding behavior of $p(x)$. 
We find 
\be 
G(x) & = & \frac{2\pi}{\sqrt{\lambda}\, x}(\Delta+ S - J) + \cdots\ , \qquad x\to \infty \nn\\ 
G(x) & = & 2\pi (m + \Mgamma  S) + \frac{2\pi x}{\sqrt{\lambda}}(S-\Delta+J) + \cdots\ , 
    \qquad x\to 0\ . 
\ee

The next step is to compare this with the usual spectral 
representation \be 
G(x) = \int_C d{x'} \frac{\sigma({x'})}{x-{x'} }\ , \qquad C = C_1 
\cup C_2 \ldots \cup C_N\ , \qquad x \in C_k\ , \ee where 
$\sigma(x)$ is a spectral density function supported on a finite 
number of cuts in the $x$ plane denoted by $C_k$.  Using the 
analyticity of $G$, we derive the following constraints \be 
\int_C d{x}\, \sigma({x}) & = & \frac{2\pi}{\sqrt{\lambda}}(\Delta+S-J)\ , \nn\\ 
\int_C d{x}\, \frac{\sigma({x})}{{x}} & = & - 2\pi (m + \Mgamma  S)\ , \nn\\ 
\int_C d{x}\, \frac{\sigma({x})}{{x}^2} & = & 
\frac{2\pi}{\sqrt{\lambda}}(\Delta-S-J)\ . \ee These may be combined 
with the condition \be p(x+i0) + p(x-i0) = 2\pi n_k\ , \qquad x \in 
C_k\ , \ee which can be understood as arising from the unimodularity 
of $\Omega(x)$. The mode numbers $n_k$ denote eigenvalues that are 
supported on the $k^{\rm th}$ contour. 
 
The conditions above yield the following finite-gap integral 
equation: \be 
&& 
    2\pi(n_k - \Mgamma J) 
    -4\pi\frac{x\, J/\sqrt{\lambda} }{x^2-1} 
    \,\, = \,\, 
\nn\\ 
&&\kern+50pt 
    2\pint_{\kern-5pt C} dx'\,\sigma(x')\left( 
    \frac{1}{x-x'} 
    -\frac{2x' + \Mgamma\sqrt{\lambda}\, ({x'}^2-1)}{2\, {x'}^2(x^2-1)} 
    + \frac{\Mgamma\sqrt{\lambda}}{2}\, \frac{1}{{x'}^2} 
    \right)\ . 
\ee Relative to the undeformed theory in the limit $\gamma \to 0$, 
we have acquired an overall shift in the mode number $n_k$ 
proportional to $\gamma$, as well as a number of $\gamma$-dependent 
modifications appearing in the integrand. In the following sections 
we will demonstrate that this thermodynamic Bethe equation can be 
promoted to a discrete Bethe equation that reliably encodes string 
energy spectra in the near-pp-wave limit.  To do this, we must first 
gather spectral data directly from the string theory.

\section{String spectra} 
\label{SPEC} 
The canonical, gauge-fixed lightcone Hamiltonian in the bosonic 
$su(2)_\gamma$ sector of the string theory in the near-pp-wave limit 
can be split into a free quadratic theory (the full pp-wave limit) 
and an interaction correction, according to \be H_{\rm LC} = H_0 + 
\frac{H_{\rm int}}{R^2} + O(1/R^4)\ . \ee Following the methods 
described in detail in Refs.~\refcite{Callan:2003xr,Callan:2004uv,Callan:2004ev,Swanson:2005wz}, we find the 
following explicit expressions in terms of worldsheet fields: \be 
H_0(S^5_{\METgamma}) 
     &=& \frac{1}{2p_-}\Bigl[ 
    4 |p_y|^2 + |y'|^2 - i p_-(y' \bar y - y \bar y')\METgamma 
   + p_-^2 |y|^2 (1+\METgamma^2) \Bigr]\ , 
\nn\\ 
{H}_{\rm int}(S^5_{\METgamma}) 
     &=& \frac{1}{8  p_-^3 }\Bigl\{ 
    -4p_y^2 (4 \bar p_y^2 + p_-^2 y^2 - {y'}^2) 
    -16p_-^2|p_y|^2|y|^2 
\nn\\ 
&&    +p_-^2 \bar y^2 (3p_-^2 y^2 + {y'}^2 - 4\bar p_y^2) 
    + \bar {y'}^2 (p_-^2 y^2 - {y'}^2 + 4\bar p_y^2) 
\nn\\ 
&&    -2 i p_-\METgamma\Bigl[ 
    -4 p_y^2 y y' + p_-^2 |y|^2(y \bar y' - \bar y y') 
    +\bar y' (4 \bar p_y^2 \bar y - {y'}^2 \bar y + y |y'|^2) 
    \Bigr] 
\nn\\ 
&&    - p_-^2 \METgamma^2 \Bigl[ 
    4 p_y^2 y^2 + \bar y^2(4 \bar p_y^2 + 2 p_-^2 y^2 - {y'}^2) 
    +4 |y|^2 |{y'}|^2 
    - y^2 \bar {y'}^2 
\nn\\ 
&&    + 2 i p_- \METgamma |y|^2 (y\bar y' - y'\bar y) 
    + p_-^2\METgamma^2 |y|^4 
    \Bigr] 
    \Bigr\}\ . 
\ee 
In truncating to the $su(2)_\gamma$ sector, we have projected onto 
the complex $y$ coordinates (though this does not achieve the complete projection). 
The free Hamiltonian can be solved and quantized exactly, yielding the 
following dispersion relations: 
\be 
\omega_n^2 = p_-^2 + (n-p_-\METgamma)^2\ , \qquad \bar\omega_n^2 = 
p_-^2 + (n+p_-\METgamma)^2\ , \ee where the integers $n$ are mode 
indices. To complete the projection to the closed $su(2)_\gamma$ 
subsector, we keep one set of bosonic raising and lowering Fourier 
modes, either $(a_n,a_{-n}^\dagger)$ or $(\bar a_n, \bar 
a_{-n}^\dagger)$. We thus choose a basis of unperturbed string 
energy eigenstates spanned by \be a_{n_1}^\dag a_{n_1}^\dag \cdots 
a_{n_N}^\dag \ket{J}\ , \ee where $\ket{J}$ is understood to be a 
ground state carrying angular momentum $J$ on the $S^5$. 
 
These states carry conserved impurity number $N$, labeled by $N$ integer mode 
numbers $n_j$.  We choose to organize these numbers such that the set 
$\{ n_j \}$ contains $M$ subsets of $N_j$ equivalent mode numbers 
$n_j$, with $j \in 1,\ldots,M$: 
\be 
\{n_j\} = 
    \Bigl\{ 
    \{ \underset{N_{1}}{\underbrace{ n_1,n_1,\ldots ,n_1 }} \}, 
    \{ \underset{N_{2}}{\underbrace{ n_2,n_2,\ldots ,n_2 }} \}, 
    \ldots , 
    \{ \underset{N_{M}}{\underbrace{ n_M,n_M,\ldots ,n_M }} \} 
    \Bigr\}\ . 
\ee 
 
With the identifications $ J = p_- R^2 $ and $p_- = 
1/\sqrt{\lambda'} $, which hold in the near-pp-wave limit, we 
arrange the large-$J$ expansion of energy eigenvalues according to 
the formula \be E(\{n_j\},J) = \sum_{j=1}^N \sqrt{1+(n_j\ 
-\METgamma/\sqrt{\lambda'})^2 \lambda'} 
    + \delta E(\{n_j\},J) + O(1/J^2)\ . 
\ee Upon diagonalizing the Hamiltonian, we find the following 
interaction correction to the free theory in the $su(2)_\gamma$ 
sector: \be 
\delta E_{\su(2)_\gamma} (\{n_j\},\{N_{j}\},J) & = & 
    -\frac{1}{2J}\biggl\{ 
    \sum_{j=1}^M N_{j} (N_{j}-1) 
    \left[\left(1+(\METgamma - n_j\sqrt{\lambda'})^{-2}\right)^{-1} \right] 
\nn\\ 
&& 
\kern-40pt 
    - \sum_{j,k = 1 \atop j \neq k}^M \frac{N_{j} N_{k}}{\omega_{n_j}\omega_{n_k}\lambda'} 
    \Bigl\{  -\lambda'(n_j n_k + n_k^2 + n_j^2(1+n_k^2 \lambda')) 
\nn\\ 
&& 
\kern-40pt 
    + \METgamma ((n_j+n_k)\sqrt{\lambda'}-\METgamma) 
    (3 + 2n_j n_k \lambda' - (n_j + n_k)\sqrt{\lambda'}\METgamma + \METgamma^2) 
\nn\\ 
&& 
\kern-40pt 
    + \lambda' (n_j\sqrt{\lambda'}-\METgamma)(n_k\sqrt{\lambda'}-\METgamma)\omega_{n_j}\omega_{n_k} 
    \Bigr\} \biggr\}\ . 
\ee In fact, this expression is identical to the one first computed 
in Ref.~\refcite{McLoughlin:2004dh}, with a global shift in the mode 
numbers $n_j \to n_j - \beta J$, where $\beta = \gamma = \tilde 
\gamma / \sqrt{\lambda}$. 
 
In the deformed $sl(2)_\gamma$ sector we find a 
near-pp-wave canonical Hamiltonian similar to the $su(2)_\gamma$ case 
described above: 
\be  
H_0(AdS_5^\METgamma) 
     &=& \frac{1}{2p_-}\Bigl[ 
    4 |p_v|^2 + |v'|^2 - i p_-(v' \bar v - v \bar v')\METgamma 
    + p_-^2 |v|^2 (1+\METgamma^2) \Bigr]\ , 
\nn\\ 
H_{\rm int}(AdS_5^\METgamma) 
     & = & \frac{1}{8p_-^3}\biggl\{ 
    16 p_-^2 |p_v|^2 |v|^2 
    + (4 \bar p_v^2 - v'^2)\bar v'^2 
    + 4 i p_-^3 |v|^2(v \bar v' - v' \bar v)\METgamma (1+\METgamma^2) 
\nn\\ 
&& 
\kern-40pt 
    + p_-^4 |v|^4(-1+6\METgamma^2+3\METgamma^4) 
    + 4 p_v^2(-4 \bar p_v^2 + v'^2 + p_-^2 v^2 (1+\METgamma^2) ) 
\nn\\ 
&&\kern-40pt 
    +p_-^2\Bigl[ 
    4 |v|^2 |v'|^2 \METgamma^2 
    + 4 \bar p_v^2 \vc^2 (1+\METgamma^2) 
    -v'^2 \vc^2(1+\METgamma^2) 
    -v^2 \vc'^2 (1+\METgamma^2) \Bigr] 
    \biggr\}\ . 
\ee  
By again solving the free limit of the theory, expanding in 
Fourier modes and projecting completely onto the closed 
$sl(2)_\gamma$ sector, we find the following correction to the 
energy spectrum at $O(1/J)$: \be 
\delta E_{\Sl(2)_\gamma} (\{n_j\},\{N_{j}\},J) & = & 
    \frac{1}{2J}\biggl\{ 
    \sum_{j=1}^M N_{j} (N_{j}-1) 
    \frac{ (\METgamma - n_j\sqrt{\lambda'})^2}{\omega_{n_j}^2\lambda'} 
\nn\\ 
&& 
\kern-80pt 
    + \sum_{j,k = 1 \atop j \neq k}^M \frac{N_{j} N_{k}}{\omega_{n_j}\omega_{n_k}\lambda'} 
    \Bigl\{ 3\METgamma^2 + \METgamma^4 - (n_j+n_k)\METgamma^3\sqrt{\lambda'} 
    + n_j n_k\lambda' (1-n_j n_k\lambda') 
\nn\\ 
&& 
\kern-40pt 
    + (n_j +n_k)\METgamma\sqrt{\lambda'}(n_j n_k \lambda'-2) 
    + \lambda'(n_k n_k\lambda' - \METgamma^2)\omega_{n_j}\omega_{n_k} 
    \Bigr\} 
    \biggr\}\ . 
\ee

\section{Bethe equations} 
\label{BETHE} 
We would now like to determine whether we can algorithmically derive 
a set of discretized Bethe equations that encode the above energy 
spectra in the near-pp-wave limit of the string theory, following the 
procedures outlined in Refs.~\refcite{Arutyunov:2004vx,Kazakov:2004qf,Beisert:2004hm,Staudacher:2004tk}. 
Generally speaking, this rests on 
the premise that the spectrum is in fact described by the 
diffractionless scattering of elementary excitations on the 
worldsheet. The excitation momenta should then obey a fundamental 
equation (see, e.g.,~Ref.~\refcite{Staudacher:2004tk}) \be p_k J = 
2\pi n_k + \sum_{j\neq k} \theta(p_k,p_j)\ , \ee so that the 
spectrum is encoded in a two-body factorized $S$ matrix 
$S(p_k,p_j)$: \be \theta(p_k,p_j) = -i \log S(p_k,p_j)\ . \ee This 
generally means that the theory also admits an infinite number of 
hidden local charges arising as linear combinations of local 
dispersion relations $q_r$: \be \label{disp} Q_r = \sum_k q_r(p_k)\ 
. \ee Adopting this language, we write the $O(1/J)$ corrections to 
the string energy spectrum in the near-pp-wave region as \be 
\delta\Delta(n_k,n_j,\Mgamma) 
    = \lambda' \sum_{j,k=1\atop j\neq k}^S 
     \frac{J}{2\pi}\frac{n_k}{\sqrt{1+\lambda' n_k^2}}\, 
    \theta\left({2\pi}n_k/J,{2\pi}n_j/J \right)\ . 
\ee

One issue arises, however, when interpreting $\sigma(x)$ (introduced 
in Section \ref{INTEG}) as a density function $\rho$ of Bethe roots 
for the string sigma model. The following integral appears with 
incorrect normalization: \be \int_C d{x}\, \sigma({x})  \sim 
\Delta+S-J\ . \ee  One way to fix this is to apply a nonlinear 
redefinition of the spectral 
parameter\cite{Kazakov:2004qf,Beisert:2004hm} $\varphi \equiv x + 
{T}/{x}$, such that $T \equiv \frac{\lambda'}{16\pi^2}$, and 
$\rho(\varphi) = \sigma(x)$. Under this change of variables, the 
thermodynamic Bethe ansatz becomes \be 2\pint d\varphi' 
\frac{\rho(\varphi')}{\varphi-\varphi'} 
    & = & 2\pi (n_k - \Mgamma J) - p(\varphi) 
\nn\\ 
&&\kern-50pt 
    + \pint d\varphi' \rho(\varphi') 
    \biggl\{ 
    \frac{2T}{\sqrt{{\varphi'}^2-4T}\sqrt{{\varphi}^2-4T}} 
    \left(\frac{x}{T-x x'} - \frac{x'}{T-x x'} \right) 
\nn\\ 
&&\kern+30pt 
    + 4\pi \Mgamma J T 
    \left( 
    \frac{1}{x^2-T} - \frac{1}{{x'}^2-T} 
    \right) 
    \biggr\}\ . 
\ee 
 
At this point, following Ref.~\refcite{Arutyunov:2004vx}, we should 
be able to recast the expression on the right-hand side in terms of 
the dispersion relations 
\be q_r(\varphi) = 
\frac{1}{\sqrt{\varphi^2-4T}} 
    \left(\frac{1}{2}\varphi+\frac{1}{2}\sqrt{\varphi^2-4T}\right)^{1-r}\ . 
\ee 
The undeformed sectors are known to arise from the geometric sum\cite{} 
\be 
    -2 \pint d\varphi' \rho(\varphi') \biggl\{ 
    \sum_{r=1}^\infty T^{r}\left( 
    q_{r+1}(\varphi')q_{r}(\varphi)-q_{r}(\varphi')q_{r+1}(\varphi) 
    \right) 
    \biggr\}\ , 
\ee while the deformation terms come from the 
combination\footnote{These expressions can be further simplified by 
rewriting the momenta in terms of the two constrained variables 
$x^\pm$ introduced in Ref.~\refcite{Beisertsu22}.  For the sake of 
exposition, however, we keep the notation originally used in 
Ref.~\refcite{US}.} \be 
    -4 \pi \Mgamma J T\, \pint d\varphi' \rho(\varphi') 
    \left( 
    q_2(\varphi)- q_2(\varphi') \right)\ . 
\ee 
We therefore obtain the following thermodynamic Bethe ansatz: 
\be 
\label{continuum} 
2\pint d\varphi' \frac{\rho(\varphi')}{\varphi-\varphi'} 
    & = & 2\pi (n_k - \Mgamma J) - p(\varphi) 
\nn\\ 
&&\kern-60pt 
    -2 \pint d\varphi' \rho(\varphi') \biggl\{ 
    \sum_{r=1}^\infty T^{r}\left( 
    q_{r+1}(\varphi')q_{r}(\varphi) \right. 
\nn\\ 
&&\kern-50pt 
   \left. 
   -q_{r}(\varphi')q_{r+1}(\varphi) 
    \right) 
    +2 \pi \Mgamma J  T \left( 
    q_2(\varphi)- q_2(\varphi') \right) 
    \biggr\}\ . 
\ee

The above result can be understood to arise from the thermodynamic 
limit of a discrete ansatz for the deformed $su(2)_\gamma$ sector of 
the theory: \be 
e^{i(p_k-2\pi \Mgamma ) J} = \prod_{j=1 \atop j \neq k}^S 
    \frac{\varphi(p_k) - \varphi(p_j) - i}{\varphi(p_k) - \varphi(p_j) + i}\, 
    e^{- 2\pi i \Mgamma g^2  ( q_2(p_k)- q_2(p_j))} 
    \prod_{r=1}^\infty e^{-2i\theta_r(p_k,p_j)}\ , 
\ee 
where $\theta_r(p_k,p_j) \equiv \left( {g^2}/{2} \right)^r \left( 
    q_r(p_k) q_{r+1}(p_j) - q_{r+1}(p_k) q_{r}(p_j) 
    \right)$. We therefore find the following $\gamma$-dependent deformation 
contribution to the worldsheet $S$ matrix in the strong-coupling 
limit: 
\be 
\theta(p_k,p_j,\Mgamma) \approx -\frac{2}{\varphi(p_k)-\varphi(p_j)} 
    -2 \sum_{r=1}^\infty \theta_r(p_k,p_j) 
    - 2\pi \Mgamma g^2 
    \left(  q_2(p_k)- q_2(p_j) \right)\ . 
\ee It is straightforward to check that this discrete Bethe ansatz 
correctly reproduces the $O(1/J)$ energy shift in the $sl(2)_\gamma$ 
sector of the string theory in the near-pp-wave limit.  The 
corresponding modifications to the discrete Bethe equations in the 
deformed $su(2)_\gamma$ subsector are comparatively simple:  they 
amount to an overall $\gamma$-dependent shift in the mode indices 
(see Ref.~\refcite{US} for further details).

\section{Conclusions} 
\label{CONC} 
The investigation summarized in this letter was intended to provide 
a number of consistency checks on the methodology proposed in the 
literature for deriving discrete Bethe equations encoding the energy 
spectra of certain sectors of type IIB superstring theory on 
$AdS_5\times S^5$. While promising results had been established 
(see, 
e.g.,~Refs.~\refcite{Arutyunov:2004vx,Kazakov:2004qf,Beisert:2004hm}), 
there was certainly no guarantee at the time that this methodology 
would work in the more complicated case of string theory on a 
TsT-deformed $AdS_5$ subspace.  It was therefore satisfying to see 
that one could indeed find Bethe equations that properly reproduced 
(in a highly nontrivial manner) the leading $1/J$ corrections to the 
energy spectrum away from the pp-wave limit. (See also 
Ref.~\refcite{Swanson:2004qa,McLoughlin:2005gj} for similar studies.) 
 
Since the publication of Ref.~\refcite{US} (on which the seminar 
reviewed here was based), the study of the (undeformed) $S$ matrix 
describing the string and gauge theory has greatly improved. A 
recent proposal by Beisert, Eden and Staudacher\cite{Beisert:2006ez} 
passes many nontrivial tests, and stands as a strong candidate for 
the complete $S$ matrix of the theory.  It would be interesting to 
consider in this larger context the types of deformations discussed 
in Ref.~\refcite{US}.  A first step would be to compute the leading 
finite-$\lambda$ corrections to the strong coupling limit of the $S$ matrix in
these deformed string theories.

\section*{Acknowledgments} 
The author would like to thank Tristan McLoughlin for collaboration 
on Ref.~\refcite{US}.  The seminar talk summarized here was based on 
this article.  I.S.~is the Marvin L.~Goldberger Member at the 
Institute for Advanced Study, and is supported additionally by 
U.S.~National Science Foundation grant PHY-0503584. 
 

\end{document}